\documentclass[12pt,preprint]{aastex}

\def\be{\begin{equation}}
\def\ee{\end{equation}}
\def\bea{\begin{eqnarray}}
\def\eea{\end{eqnarray}}

\usepackage{graphicx}

\begin{document}

\title{Optical/infrared flares of GRB 080129 from late internal shocks}

\author{Wei-Hong Gao$^1$}

\affil{$^1$ Department of Physics and Institute of Theoretical
Physics,
Nanjing Normal University, Nanjing, 210097, China\\
}

\email{gaoweihong@njnu.edu.cn}

\begin{abstract}
Strong optical and near-infrared (NIR) flares were discovered in the
afterglow of GRB 080129. Their temporal behaviors, the sudden
emergence and the quick disappearance, are rather similar to that of
many X-ray flares (for instance, the giant flare of GRB 050502B). We
argue that the optical/NIR flares following GRB 080129 are a low
energy analogy of the X-ray flares and the most likely
interpretation is the ``late internal shock model". In this model,
both the very sharp decline and the very small ratio between the
duration and the occurrence time of the optical/NIR flares in GRB
080129 can be naturally interpreted. The initial Lorentz factor of
the flare outflow is found to be $\sim 30$, consistent with the
constraint $\leq 120$ set by the forward shock afterglow modeling.
Other possibilities, like the reverse shock emission or the
radiation from the continued but weaker and weaker collision between
the initial GRB outflow material, are disfavored.
\end{abstract}


\section{INTRODUCTION}
GRB 080129 was triggered and located by the {\it Swift} Burst Alert
Telescope (BAT) at 06:06:45 UT \cite{Immler08}. The duration of
prompt emission $T_{90}$ is $48\pm 10$ sec in $15-350$ keV band
\cite{Barthelmy08}. The time-averaged spectrum is best fitted by a
simple power-law model, whose power law index is $1.34 \pm0.26$. The
fluence in the $15-150$ keV band is $8.9 \pm 1.4 \times 10^{-7} \rm
erg/cm^{2}$ \cite{Barthelmy08}. The BAT observations lasted until
320 seconds after the trigger, then slewed to another location of
the sky. X-ray telescope(XRT) and the UV-optical telescope(UVOT)
started to point to GRB 080129 until $3.2\times 10^{3}$s after the
trigger. A fading X-ray source was discovered and no emission was
seen with UVOT. No flare was observed in X-ray band since XRT
started to observe \cite{Holland08}.

The optical/NIR observations imaged by GROND started immediately
after the trigger \cite{Greiner08}. The first images immediately
revealed a strongly flaring source. Distinguished optical/NIR flares
were observed with amplitude $\sim 3 $ mag, duration of 80 s
(full-width at half maximum; FWHM, hereafter we define the FWHM as
the observed variability timescale $\delta t$ in the flare), peaking
at $t_{\rm p}\sim 540 $ s after the GRB trigger. Their rise and the
decline can be well approximated by $t^{12}$ and $t^{-8}$,
respectively. Thereafter, the afterglow brightness is continuously
rising until 6000 s after the GRB. The optical spectroscopy suggests
a redshift $z=4.349$ for GRB 080129.

\begin{figure}
\plotone{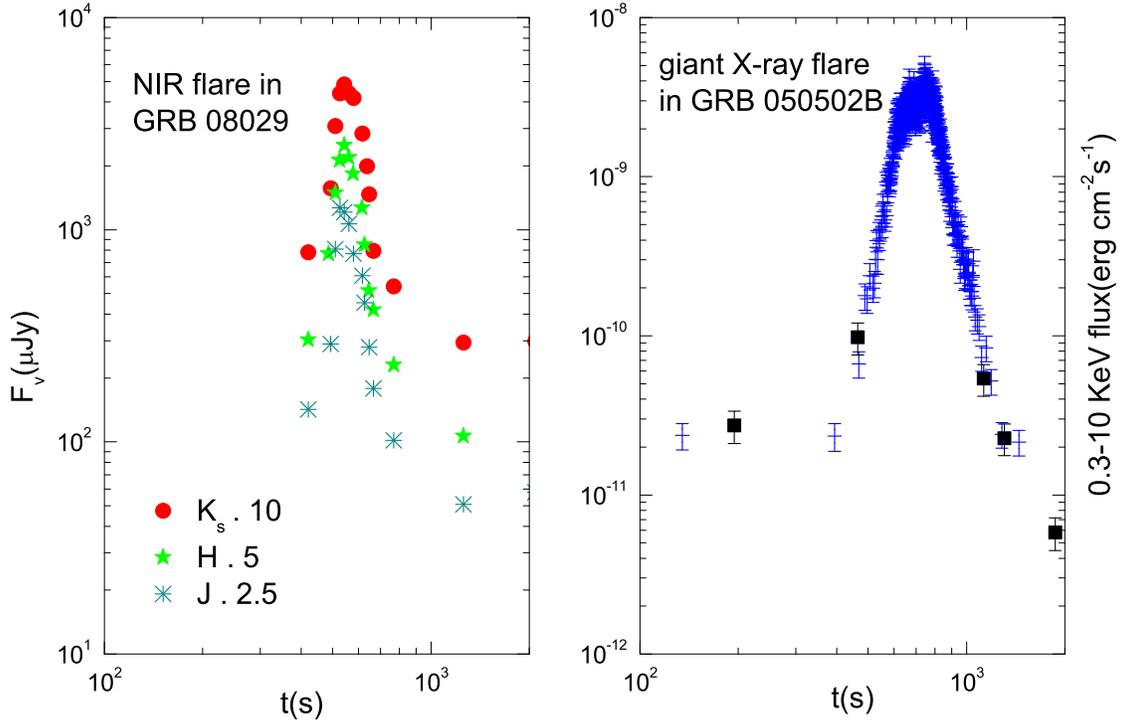} \caption{The light curves of NIR flares in GRB
080129 compare with that in GRB 050502B in X-ray band. Left: The red
circles, green pentagons and blue points represent observed data
from three near-infrared channels $\rm K_{s}\times$10, H$\times$5,
J$\times$2.5 on GROND, respectively\cite{Greiner08}. Right: Data get
from UNLV GRB Group(http://grb.physics.unlv.edu/). {\bf The main
parameters describing these flares are summarized in
Tab.\ref{tab:1}.} }\label{fig:1}
\end{figure}

Greiner et al.(2008) interpreted the optical/NIR flares as the
radiation of continued but weaker and weaker collisions between the
material ejected during the prompt emission phase. {\it In this work
we do not follow their treatment for the following arguments:} (i)
In such a scenario, the NIR/optical flares emerge when the
synchrotron self-absorption frequency drops below the observer's
frequencies. If correct, the NIR and optical flares should have an
observable/significant time delay, that is the higher the observer's
frequency, the earlier the arrival time. However, we did not see
such a delay in the data (Greiner et al. 2008; see also our
Fig.\ref{fig:1}). (ii) The NIR/optical flares appeared and then
peaked at a time $t\sim 540 {\rm s}\gg T_{90}$. If the IR/optical
flares are indeed from the outflow material ejected during the
prompt gamma-ray emission phase, their declines are governed by the
high latitude emission and can not be steeper than $(t-T_{\rm
90})^{-(2+\beta)}\approx t^{-(2+\beta)}$, again inconsistent with
the data, where $\beta\leq p/2$ is the spectral index, $p$ is the
power-law index of the energy distribution of the shock-accelerated
electrons \citep{Kumar00,Fan05}. This puzzle can be solved if the
jet is so narrow that we have seen its edge, i.e., $\theta_{\rm j}
\leq 0.01~(100/\Gamma_{\rm i})$, where $\theta_{\rm j}$ is the half
opening angle and $\Gamma_{\rm i}$ is the initial Lorentz factor of
the outflow. However such a possibility has been convincingly ruled
out by the late time afterglow observation because the jet break at
$\sim 1.8\times 10^{4}$ s suggests a $\theta_{\rm j} \sim 0.076 \gg
1/\Gamma_{\rm i}$ \cite{Greiner08}. The latter argument applies to
the reverse shock emission model as well. That's why we won't
discuss such a possibility in this work, either.

We note that the temporal behavior of the NIR/optical flares
detected in GRB 080129 is quite similar to that of X-ray flares
observed in a good fraction of Swift GRB afterglows (e.g., Guetta et
al. 2006; Chincarini et al. 2007). For comparison purpose,  we
re-plot both the giant X-ray flare following GRB 050502B
\citep{Burrows05} and the NIR flares in GRB 080129 in
Fig.\ref{fig:1}. The physical parameters are summarized in
Tab.\ref{tab:1} and the similarities are evident. {\it Motivated by
these similarities we suggest that the NIR flares detected in GRB
080129 should have the same origin of the flares observed in the
X-ray afterglows, i.e., the NIR flares should be powered by the
so-called late internal shocks, too}
\citep{Fan05,Burrows05,Zhang06}. Such a model was thought to have
been ruled out by the request of a very large initial Lorentz factor
($\Gamma_{\rm i}\sim 800$) of the flare outflow \citep{Greiner08}.
We'll show in this work that $\Gamma_{\rm i}\sim $ tens, a typical
value taken in the X-ray flare modeling \citep{Fan05}, is large
enough to reproduce the data and is consistent with the upper limit
($\leq 120$) set by the forward shock optical afterglow modelling of
GRB 080129.

\begin{table}[ht]
\caption[SED]{The main parameters governing the opical/NIR flares in
GRB 080129 and the X-ray flare in GRB 050502B.

 \label{tab:1}}
  \bigskip
  {\small
  \begin{tabular}{ccccc}
  \hline
  \noalign{\smallskip}
 flare(s) in GRB  &   rise index ($\alpha_1$) & decline index ($-\alpha_2$) & $t_{\rm p}$ (s) & $\delta t/t_{\rm p}$ \\
  \noalign{\smallskip}
\hline
  \noalign{\smallskip}
GRB 080129 & 12& $-8$ & 540 &  0.15  \\
GRB 050502B & 9.5 & $-9.0$ & 740 & 0.14 \\
  \noalign{\smallskip}
  \hline
  \end{tabular}}
\end{table}

This work is arranged as the following. In section 2 we briefly
introduce the late internal shock model and then discuss the
identity of two independent constraints that are widely used to rule
out other possibilities. In section 3 we apply the late internal
shock model to the IR flares following GRB 080129. We summarize our
results with some discussions in section 4.

\section{THE LATE INTERNAL SHOCK MODEL}
In the standard fireball model, the GRB prompt emission is powered
by the interaction of shells with different Lorentz factors in the
relativistic outflow launched by the central engine, i.e., the
internal shock model \cite{Paczynski94,rees94} while the afterglow
is believed to be the external forward shock emission
\cite{Piran99}. However, since the launch of {\it Swift} satellite,
energetic X-ray flares have been detected in about half of the GRBs
afterglows. The temporal behavior of most X-ray flares share some
similarities with the prompt soft $\gamma-$ray emission and can not
be interpreted by the external forward shock model (See M\'esz\'aros
2006; Zhang 2007 for recent reviews). The most likely interpretation
is the so-called ``late internal shocks model", in which the GRB
central engine restarts after the prompt emission phase and launches
unsteady outflow. The underlying physical processes are less clear.
Among the various models put forward (see Zhang 2007 for a review)
fallback accretion onto the nascent black hole may be the most
natural one. The collision between the fast and slow material of the
new outflow can power strong flares peaking in X-ray or
far-ultraviolet band. The duration of these flares ($\delta t$) is
determined by the re-activity process of the central engine and can
be much shorter than the occurrence time of the flares. On the other
hand, since the ejection time $(\sim t_{\rm eje})$ of the last main
pulse of the flare is close to $t_{\rm p}$, the net flux of the high
latitude emission of the pulses can be approximated by $(t_{\rm
p}-t_{\rm eje})^{-(2+\beta)}$, which can be much steeper than
$t^{-(2+\beta)}$. So the late internal shock model can naturally
account for the main characters, the sudden emergence and then a
rapid drop, of the X-ray flares detected so far. For the emission of
the external shocks, it is well known that (1) $\delta t/t$ has to
be in order of $1$ or larger \cite{NP03}; (2) the decline can not be
steeper than $t^{-(2+\beta)}$ unless the edge of the GRB ejecta is
visible. This is because the GRB outflow is curving and emission
from high latitude (relative to the observer) will reach us at later
times and give rise to a decline shallower than $t^{-(2+\beta)}$
\cite{Fenimore96,Kumar00}. Usually these two limitations have been
taken as independent evidences for the late internal shock model
(e.g., Chincarini et al. 2007). Below we show that {\it they are
highly relevant and even identical.}\footnote{Fan et al. (2008a)
pointed out this in a proceeding paper but did not prove it.}

\begin{figure}
\plotone{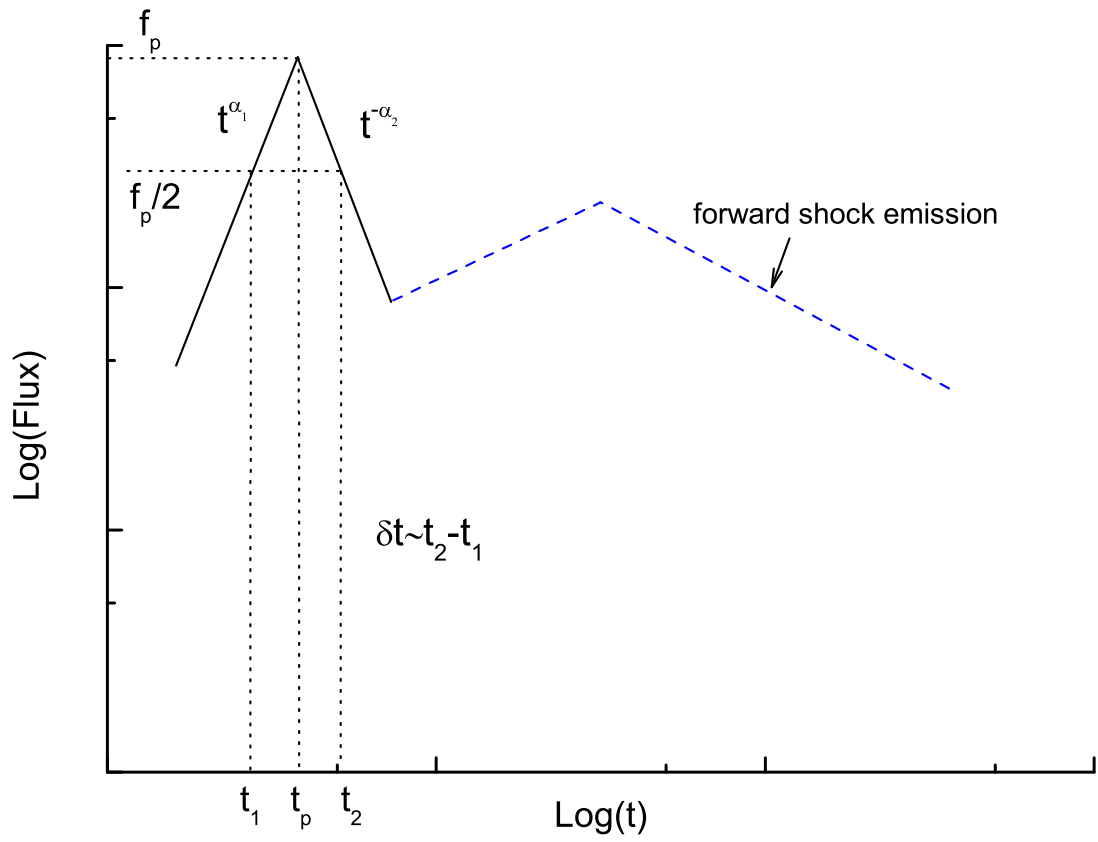} \caption{A schematic plot of a flare in the GRB
afterglow.}\label{fig:2}
\end{figure}

\begin{figure}
\plotone{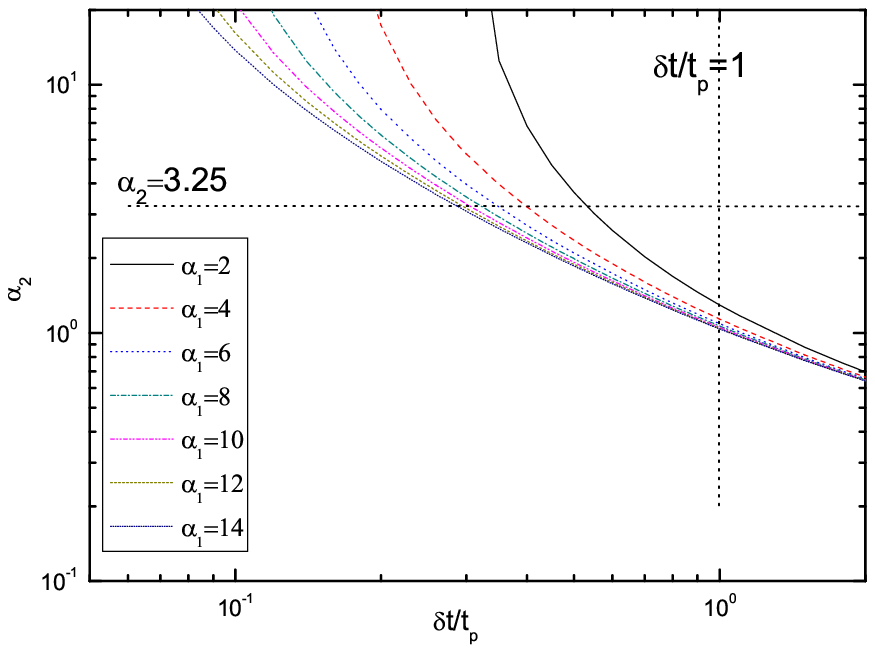} \caption{The relationship of decay power law
index $\alpha_{2}$ and the ratio of $\delta t/t_{p}$. The rising
power law index $\alpha_{1}$ is taken as 2, 4, 6, 8, 10, 12, 14,
respectively. The horizontal dot line represents
$\alpha_{2}=2+\beta=3.25$, where we take $\beta$=p/2=1.25, the
vertical dot line represents $\delta t/t_{p}=1$.}\label{fig:3}
\end{figure}

As shown in Fig.\ref{fig:2}, $f_{\rm p}$ is the maximum flux at the
peaking time $t_{\rm p}$ in the flare, $t_{1}$ and $t_{2}$ are the
time at which the flux is half of $f_{\rm p}$ in the rising and
decaying light curves, respectively. The FWHM time is therefore
$\delta t \equiv t_{2}-t_{1}$. Before and after $t_{\rm p}$, the
light curves are approximated by $t^{\alpha_1}$ and $t^{-\alpha_2}$,
respectively. It is straightforward to see that $\rm
\alpha_{1}=\frac{Log(f_{p})-Log(\frac{1}{2}f_{p})}{Log(t_{p})-Log(t_{1})}$,
and $\rm
\alpha_{2}=\frac{Log(f_{p})-Log(\frac{1}{2}f_{p})}{Log(t_{2})-Log(t_{p})}$.
After some simple algebraic we have
\begin{eqnarray}
\delta t/t_{\rm p}=2^{1/\alpha_{2}}-2^{-(1/\alpha_{1})} \label{Eq:1}
\end{eqnarray}
Obviously, $\delta t/t_{\rm p}$ is irrelevant to $f_{\rm p}$. For
fixing $\alpha_{1}$, the relationship between $\delta t/t_{\rm p}$
and $\alpha_{2}$ can be found in Fig.\ref{fig:3}, with which we can
see that the steeper the decay, the smaller the $\delta t/t_{\rm
p}$. Particularly for $\alpha_1\gg 1$, we have $\alpha_2\gg 2+\beta$
(hereafter ``the decline constraint") and $\delta t/t_{\rm p}\ll 1$
simultaneously, suggesting that the two constraints widely used in
supporting late internal shock model are highly relevant. This
naturally accounts for the fact that many X-ray flares satisfy both
limitations \cite{Chincarini07}. Please note that our conclusion is
independent of the underlying physical processes.

For the purpose of identifying the afterglow emission powered by the
central engine, the decline constraint may be more general. For
example, the very sharp drop detected in the X-ray afterglow of GRBs
070110 \citep{Troja07}, 060413, 060522, 060607A and 080330
\citep{Zhang09} also favors a central engine origin though the
constraint $\delta t/t\ll 1$ is violated.

For the optical/NIR flares of GRB 080129, the peak time is $\sim
540$ s after the trigger, the duration (i.e., the FWHM) is $80$ s.
We have $\delta t/t_{\rm p}\sim 0.15\ll1$ and $\alpha_2= 8$,
indicating a central engine origin of these flare photons. Below we
take the late internal shock model to reproduce the data.

\section{\rm Physical parameters and the synchrotron radiation of
GRB 080129 in late internal shock model} We assume that the Lorentz
factors of the ejected material in the re-starting outflow are
highly variable, and take $\rm \Gamma_{s}\sim 10$ and $\rm
\Gamma_{f}\sim 100$ as the typical Lorentz factor of the slow and
fast shells, respectively. The masses of the fast and slow shells
are taken as $\rm m_{f} \simeq m_{s}$. In the late internal shock
model, the inner fast shell will catch up with the outer slow shell
at the radius $\rm \sim 2\Gamma_{s}^{2}c \delta t_{i}/(1+z)$(where
$\rm \delta t_{i}$ is taken as the observed typical variability
timescale of one pulse in GRB 080129 optical/NIR flare), and
internal shock are generated. The merged shell's Lorentz factor is
$\Gamma_{\rm i}\approx\rm \sqrt{\Gamma_{f}\Gamma_{s}}\sim 30$
\citep{Piran99}, and the Lorentz factor of the internal shock can be
estimated as $\rm
\gamma_{sh}\approx(\sqrt{\Gamma_{f}/\Gamma_{s}}+\sqrt{\Gamma_{s}/\Gamma_{f}})/2$.

%Throughout the paper we follow the convention $\rm F_{\nu,t}\propto
%\nu^{-\beta}t^{\alpha}$.
Adopting the cosmological parameters $\rm H_{0}=70 km s^{-1}
Mpc^{-1}$, $\Omega_{\rm M}=0.3$, and
 $\Omega_{\Lambda}=0.7$, we have a luminosity distance $\rm
D_{L}=1.2\times10^{29}$ cm for GRB 080129 at a redshift $z=4.35$.
The observed maximum flux of the flare in GRB 080129 is about$\sim$
0.5 mJy (NIR) and $\sim$0.3 mJy (optical), respectively
\cite{Greiner08}. Assuming an efficiency factor of the optical flare
$\epsilon \sim$ 0.1, the total luminosity of the flare outflow ican
be estimated by $\rm L_{m}\sim 3\times 10^{48}~{\rm erg s^{-1}}$.
This luminosity implies that the fallback accretion rate is about
$\sim 10^{-5}-10^{-3}$ times that of the GRB prompt accretion, if
the efficiency factor of converting the accretion energy into the
kinetic energy of the outflow is nearly a constant
\cite{MacFadyen01}.

The variability timescale $\delta t_{\rm i}$ of GRB080129's optical
flare is significantly longer than that of the prompt emission. Here
we take $\delta t_{\rm i}\sim 30\rm s$, as suggested by the
smoothness of the flare light curves. The typical radius of the late
internal shock is $R_{\rm int}\approx 2\Gamma_{\rm i}^{2}c\delta
t_{\rm i}/(1+z)\approx3.4\times10^{14}~\Gamma_{\rm i,1.5}^{2}\delta
t_{\rm i,1.5}~{\rm cm}$. In this work we take the convenience
$Q_{\rm x}=Q/10^{\rm x}$ in units of cgs unless with specific
notation. Below, following Fan \& Wei (2005), we show that with the
parameters $\Gamma_{\rm i} \sim 30$, $L_{\rm m}\simeq3\times 10^{48}
\rm erg~{\rm s}^{-1}$, $\epsilon_{\rm e}=0.4$, $\epsilon_{\rm
B}=0.05$, and $\delta t_{\rm i}=30 \rm s$, the flare data can be
reasonably reproduced.

We first investigate the interaction between the inner fast shell
and the outer slower one. The comoving number density of the
electrons is $n_{\rm e}\simeq L_{\rm m}/(4\pi \Gamma_{\rm i}^2 R^2
m_{\rm p}c^3)\simeq 4.6\times 10^{7}L_{\rm m,48.5}\Gamma_{\rm i,
1.5}^{-6}\delta t_{\rm i,1.5}^{-2}$ , where $m_{\rm p}$ is the rest
mass of proton. The thermal energy density of the shocked material
is $e=4\gamma_{\rm sh}(\gamma_{\rm sh}-1)n_{\rm e}m_{\rm p}c^{2}$
\cite{Blandford76}. So  the strength of magnetic field can be
estimated as
\begin{eqnarray}
\rm B &\approx& (8\pi \epsilon_{B}e)^{1/2}\approx 6.7\times
10^{2}~{\rm
G}~\rm \epsilon_{B,-1.3}^{1/2}(\frac{\rm \gamma_{sh}}{2})^{1/2}\nonumber \\
&&\rm (\gamma_{sh}-1)^{1/2}L_{m,48.5}^{1/2}\Gamma_{\rm i,
1.5}^{-3}\rm \delta t_{\rm i,1.5}^{-1}.
\end{eqnarray}

As usual, we assume that in the shock front, the accelerated
electrons take an energy distribution $dn_{\rm e}/d\gamma_{\rm
e}\propto \gamma_{\rm e}^{-p}$ for $\gamma_{\rm e}>\gamma_{\rm
e,m}$, where $\gamma_{\rm e,m}=\epsilon_{\rm e}(\gamma_{\rm
sh}-1)[(p-2)m_{\rm p}/(p-1)m_{\rm e}]$ is the minimum Lorentz factor
of the shocked electrons \cite{Sari98}, and $m_{\rm e}$ is the rest
mass of an electron. Here for GRB 080129, we take $p=2.5$. We can
get the observed typical frequency of the synchrotron radiation

\begin{eqnarray}
\nu_{\rm m} &=& \gamma_{\rm e,m}^{2}q_{\rm e}\Gamma B/[2(1+z)\pi
m_{e}c] \nonumber \\ &\simeq & 3.7\times10^{14}~{\rm
Hz}~\epsilon_{\rm e,-0.4}^{2}\epsilon_{\rm B,-1.3}^{1/2}(\gamma_{\rm
sh}-1)^{5/2}(\gamma_{\rm sh}/2)^{1/2} \nonumber
\\&&L_{\rm m,48.5}^{1/2}\Gamma_{\rm i,1.5}^{-2}\delta
t_{\rm i,1.5}^{-1},
\end{eqnarray}
where $q_{\rm e}$ is the charge of the electron.

The cooling Lorentz factor is estimated by $\gamma_{\rm e,c}\approx
7.7\times 10^{8}(\rm 1+z)/(\rm \Gamma B^{2}\delta t_{\rm i})$. So
the cooling frequency is \cite{Sari98}
\begin{eqnarray}
\nu_{\rm c}&=&\gamma_{\rm e,c}^{2}q_{\rm e}\rm \Gamma B/[2(1+z)\pi
\rm m_{\rm e}c]\nonumber
\\ &\rm \simeq& 1.0\times10^{12}~{\rm
Hz}~\rm
\epsilon_{B,-1.3}^{-3/2}(\frac{\gamma_{sh}}{2})^{-3/2}\nonumber
\\
&&(\gamma_{\rm sh}-1)^{-3/2}L_{\rm m,48.5}^{-3/2}\Gamma_{\rm
i,1.5}^{8}\delta t_{\rm i,1.5}.
\end{eqnarray}

The synchrotron self-absorption frequency can be estimated as

\begin{eqnarray}
\rm \nu_{a} &&\approx 1.2\times10^{14}~{\rm Hz}~\rm
\epsilon_{B,-1.3}^{1/14}L_{m,48.5}^{5/14}[(\gamma_{sh}-1)\frac{\gamma_{sh}}{2}]^{1/14}\nonumber
\\ &&\Gamma_{\rm i,1.5}^{-8/7}\delta t_{\rm i,1.5}^{-5/7}.
\end{eqnarray}

The maximum spectral flux of the synchrotron radiation is $F_{\rm
max}\approx3\sqrt{3}\Phi_{\rm p}(1+z)N_{\rm e}m_{\rm
e}c^{2}\sigma_{T}\Gamma B/(32\pi^{2}q_{\rm e}D_{\rm L}^{2})$, where
$N_{\rm e}$ is the total number of emitting electrons, $N_{\rm
e}=L_{\rm m}\delta t_{\rm i}/ [(1+z)\Gamma_{\rm i} m_{\rm p}
c^{2}]=9.3\times10^{50}L_{\rm m,48.5}\Gamma_{\rm i,1.5}^{-1}\delta
t_{\rm i,1.5}$, where $\delta t_{\rm i}$ is the observed typical
variability timescale of the total flare, and $\Phi_{\rm p}$ is a
function of $p$. For $p=2.5$ we have $\Phi_{p}=0.6$ \cite{Wijers99}.
For $\nu_{c}<\nu_{a}<\nu<\nu_{m}$, the predicted flux is
\cite{Sari98}
\begin{eqnarray}
\rm F_{\nu}&=&\rm F_{max}(\nu/\nu_{c})^{-1/2}\nonumber
\\ &\sim& 5.0\times10^{-4}~\rm
Jy~[\nu/(3.0\times10^{14}~\rm Hz~)]^{-1/2}\rm
\epsilon_{B,-1.3}^{-1/4}(\frac{\gamma_{sh}}{2})^{-1/4}\nonumber
\\ &&\rm (\gamma_{sh}-1)^{-1/4}L_{m,48.5}^{3/4}\delta t_{\rm i,1.5}^{1/2}\Gamma_{\rm
i,1.5}D_{L,29}^{-2}\label{Eq:7}
\end{eqnarray}
Taking $\nu_{\rm NIR}=3.0\times10^{14}~\rm Hz~$, we have
$F_{\nu_{\rm NIR}}\sim  0.5~\rm~mJy$, consistent with the
observation of GRB 080129's flare in near-infrared band.

In the optical and X-ray band satisfying $\rm
\nu_{c}<\nu_{a}<\nu_{m}<\nu$, the flux can be estimated as $\rm
F_{\nu}= F_{max}(\nu_{m}/\nu_{c})^{-1/2}(\nu/\nu_{m})^{-p/2}$
\cite{Sari98}. Taking $\nu_{\rm opt}=5.0\times10^{14}\rm Hz$ for
optical band and $2\times 10^{17}\rm Hz$ for X-ray band, we have
$F_{\nu_{\rm opt}}\sim 0.3\rm mJy$ and $F_{\nu_{X}}\sim 1.7\times
10^{-4}\rm mJy$, respectively. Approximately the optical peak flux
of the flare is $\sim0.3 \rm mJy$, as inferred from Fig.1 of Greiner
et al.(2008). So our result is consistent with the optical data,
too. In the X-ray band, no observation was carried out for $t\leq
3.2\times10^{3}$ s. So it is impossible to test our predication in
X-ray band.

\section{DISCUSSION}
In $\sim 10^{2}-10^{5}$ s after the trigger of GRBs, bright X-ray
flares have been well detected in a good fraction of {\it Swift} GRB
X-ray afterglows (Falcone et al. 2007; Chincarini et al. 2007).
However, for many X-ray flares the peak energy is unknown and the
upper limit is about 0.2 keV. Fan \& Piran (2006) speculated that
some X-ray flares actually peaked in UV/optical band and thus should
be classified as UV/optical flares. However, before 2008 people had
not detected a canonical optical flare with plenty of data. The best
candidate of UV/optical flare may be that detected in GRB 050904
(Bo\"er et al. 2006), for which, unfortunately, the reverse shock
model can not be ruled out (Wei, Yan \& Fan 2006). The situation
changed dramatically after the release of the early optical/NIR
afterglow data of GRB 080129 \cite{Greiner08}.

The optical/NIR flares following GRB 080129 have very sharp decline
($\alpha_2 \gg 2+\beta$) and very small $\delta t/t_{\rm p}(\sim
0.15)$, rather similar to that of the giant X-ray flare following
GRB 050502B. These two characters rule out the possibility of being
the reverse shock emission or being the radiation of the continued
but weaker and weaker collision between the outflow material ejected
during the prompt emission phase. Instead, these optical/NIR flares
can be attributed to the re-activity of the central engine, as the
X-ray flares detected in a good fraction of Swift GRB X-ray
afterglows. In the framework of late internal shock model, with
reasonable physical parameters (in particular $\Gamma_{\rm i}\sim$
tens) we calculate the synchrotron radiation. The typical frequency
is just in near infrared band and the flux estimated in
near-infrared and optical band are also consistent with the
observations (see section 3 for details). We conclude that the
flares in GRB 080129 peaking in NIR/optical band are a low energy
analogy of the X-ray flares, confirming the speculation of Fan \&
Piran (2006).

The identification of a low energy analogy of X-ray flares in
optical/IR band also helps the people to diagnose the physical
composition of the outflow launched by the re-activity of the
central engine. Fan et al. (2008b) showed that polarimetry of the
flares is highly needed to achieve such a goal. Technically the
optical polarimetry is much more plausible than the X-ray
polarimetry at present \citep{Covino99}.

In this work we also show that the two constraints $\alpha_2\gg
2+\beta$ (i.e., the decline constraint) and $\delta t/t\ll 1$,
widely/separately used to support the ``central engine origin" of
the afterglow emission, are highly relevant and even identical (see
section 2 for details) for the flares. The decline constraint may be
more general. For example, the very sharp drop detected in the X-ray
afterglow of GRBs 070110 \citep{Troja07}, 060413, 060522, 060607A
and 080330 \citep{Zhang09} is in support of a central engine origin
though the constraint $\delta t/t\ll 1$ is unsatisfied.

\section*{Acknowledgments}
We thank the referee for helpful suggestions and Dr. Yizhong Fan for
the stimulating discussion and for his help on improving the
presentation. This work is supported by the National Natural Science
Foundation (grant 10603003) of China.

%%REFERENCES

\end{document}